\documentclass[runningheads]{llncs}
\usepackage{amsmath, amssymb}
\usepackage{xspace}
\usepackage[T1]{fontenc}
\usepackage{xcolor}
\usepackage{verbatim}
\usepackage{tikz}
\usetikzlibrary{decorations.pathreplacing}
\usepackage{todonotes}

\usepackage{graphicx}
\usepackage{hyperref}
\spnewtheorem*{mainrestated}
  {Theorem~\ref{thm:clusterruntime} }
 {\normalfont\bfseries}{\normalfont}

 \spnewtheorem*{maintworestated}
  {Theorem~\ref{thm:cliqueruntime} }
 {\normalfont\bfseries}{\normalfont}

\begin{document}
\title{Parameterized Algorithms for Minimum Sum Vertex Cover on Dense Graphs}
%
%
\author{Shubhada Aute\inst{1}\orcidID{0009-0000-2964-0368} \and
Noleen K\"ohler\inst{2}\orcidID{0000-0002-1023-6530} \and
Fahad Panolan\inst{2}\orcidID{0000-0001-6213-8687}}
\authorrunning{S. Aute et al.}
%
\institute{Indian Institute of Technology Hyderabad, Sangareddy, India 
\email{cs21resch11001@iith.ac.in}
\and
University of Leeds, Leeds, UK\\
\email{\{N.Koehler,F.Panolan\}@leeds.ac.uk}}
\maketitle              
\begin{abstract}

The \textsc{Minimum Sum Vertex Cover} (MSVC) problem asks for a linear ordering of the vertices of a graph that minimizes the sum, over all edges, of the position of the earlier endpoint. We study its parameterized complexity on graphs that become a clique, cluster graph, or a split graph after deleting a small set of vertices, called the Modulator.
We prove that MSVC is fixed-parameter tractable when parameterized jointly by the cluster modulator size and the maximum clique size in the remaining cluster graph, while allowing arbitrarily many cliques. For the clique modulator parameterization, we present a combinatorial algorithm that avoids the integer quadratic programming used in prior work [Aute, Panolan, Theoretical Computer Science, 2025].
We also introduce a vertex-weighted formulation of MSVC and obtain an XP algorithm parameterized by the clique modulator size. Using this algorithm as a subroutine, we develop an XP algorithm parameterized by the split modulator size. These results extend the known structural parameterizations of MSVC and highlight weighted MSVC as a promising direction for further research.

\keywords{FPT  \and Split Graphs \and Structural parameters}
\end{abstract}

\section{Introduction}
\label{sec:intro}
Vertex ordering problems seek a linear ordering of the vertices of a graph that optimizes a given objective function. Although a graph has no inherent vertex order, the order in which its vertices are processed can affect the efficiency of an algorithm. One such problem, motivated by the design of efficient algorithms for semidefinite programming \cite{burer2001projected}, is \textsc{Minimum Sum Vertex Cover} (MSVC). In this problem, an edge is covered when its first endpoint is processed, and the objective is to minimize the sum of the cover times of all edges.
\begin{definition}
Let $G$ be an $n$-vertex graph, and let $\phi : V(G) \to [n]$ be a bijection. 
The weight (or processing time or cost) of an edge $e=uv$ is $ w(e, \phi ) = \min \{\phi(u), \phi(v) \}$. The cost of the ordering $\phi$ for the graph $G$ is defined as, 
\begin{equation*}
\mu_G(\phi) = \sum_{e \in E(G)} w(e, \phi).   
\end{equation*} 
The objective of \textsc{Minimum Sum Vertex Cover} (MSVC) problem is to find an ordering $\phi$ that minimizes $\mu_G(\phi)$ over all possible orderings. 
\end{definition}

As with many vertex ordering problems, solving MSVC is NP-hard \cite{feige2004approximating}.
In an optimal ordering, the vertices appearing up to the maximum edge-covering position form a vertex cover, though not necessarily a minimum size one. Since every edge is already covered at that point, the ordering of the remaining vertices is irrelevant. Thus, MSVC can be decomposed into two problems: identifying a vertex cover induced by an optimal ordering and, once this set is known, determining its optimal internal ordering. However, finding such a vertex cover is NP-hard, and even when the cover is given, computing its optimal ordering remains NP-hard \cite{liu2025minimum}. These inherent challenges have inspired research in the field of approximation and parameterized algorithms.

A generalization of MSVC on hypergraphs is called the \textsc{Minimum Sum Set Cover} (MSSC) problem. Both MSVC and MSSC were first studied by Feige et al. \cite{feige2004approximating}.
They gave an approximation algorithm for solving MSVC and MSSC. 
It instigated a series of papers on improving the approximation factor for the problems, including lower bound for inapproximability. 
MSVC was studied for some special classes of graphs as well. 
The first work on MSVC from a parameterized complexity perspective was initiated by Aute and Panolan \cite{aute2025parameterized}. 
They considered the sizes of the vertex cover and the clique modulator (the complement of the vertex cover) as parameters and proved that MSVC is fixed-parameter tractable with respect to both parameters individually. 
A follow-up paper by Liu et al. gave a kernelization and FPT algorithm with the parameter maximum cover time of an edge \cite{liu2025minimum}. 

The time complexity scales sharply when the graphs contain a large clique or highly fragmented small clusters. Except for the parameter clique modulator, the previously studied parameters do not perform well in finding the optimal ordering for MSVC in such graphs. Hence, there is a need for better parameters to solve MSVC for these graphs. 
In this paper, we study the MSVC problem on graphs that become cluster graphs, cliques, or split graphs after the removal of a small subset of vertices.
We present FPT and XP algorithms for solving MSVC on the mentioned graphs. 

\smallskip
\noindent
{\bf Parameterized Complexity Results.}
We first consider graphs with a cluster modulator of size $k$, whose removal leaves a disjoint union of cliques, also known as a cluster graph. Let $d$ denote the maximum clique size in the graph. We allow the number of clusters to be arbitrarily large and establish the following result. 
\begin{theorem}\label{thm:clusterruntime}
    Let $M$ be a cluster modulator of size $k$ such that every clique of $G-M$ has size at most $d$. Then
    MSVC is fixed-parameter tractable parameterized by $k+d$. 
\end{theorem}
Parameterized by the clique modulator size, MSVC is already known to be FPT \cite{aute2025parameterized}. We improve its running time.

\begin{theorem}\label{thm:cliqueruntime}
    There is an FPT algorithm for MSVC  parameterized by the clique modulator size $k$, with running time $2^{O(k2^k)}n^{O(1)}$.
\end{theorem}
We also study a vertex-weighted variant of MSVC with clique modulator size as a parameter. We use WMSVC for \textsc{Weighted Minimum Sum vertex Cover} problem. 
\begin{theorem}
    There is an XP-time algorithm for WMSVC parameterized by the minimum size of a clique modulator. 
\end{theorem}

A split graph admits a partition of its vertex set into two sets, such that one part induces a clique and the other part an independent set, and MSVC is polynomial-time solvable on this class \cite{rasmussen2006efficient}. We extend this setting by allowing a bounded number of modulator vertices.

\begin{theorem}
     There is an XP-time algorithm for MSVC with the minimum split modulator size as the parameter.
\end{theorem}

\noindent
{\bf Related Work.}
The foundational work for the \textsc{Minimum Sum Vertex Cover}, and its generalization \textsc{Minimum Sum Set Cover}, was done by Feige, Lovász, and Tetali. They proved that MSVC can be approximated within a factor of $2$ using Linear Programming \cite{feige2004approximating}. 
The $ 2$-factor barrier was broken by Barenholz et al. by modifying the rounding technique in the linear program used to solve MSVC. This method improved the approximation factor to $1.99995$ \cite{barenholz2006improved}.
A significant improvement in the approximation factor was achieved through the \textsc{Generalized Minimum Sum Set Cover problem},  abbreviated as GMSSC. Input to GMSSC is a hypergraph $H$, and there is a covering requirement $k_e$ for each hyperedge $e$, such that $k_e \leq |e|$. The cover time of a hyperedge $e$ is the location of its $k_e^{th}$ endpoint in the ordering of the vertices. 
Using semi-definite programming for GMSSC, an approximation ratio of $16/9$ was achieved for MSVC, and it is the current best ratio given by Bansal et. al. \cite{bansal2021improved}. On the hardness side, Stanković proved that approximating MSVC within a ratio of $1.0748$ is NP-hard under the Unique Games Conjecture (UGC) \cite{stankovic2025some}.

Although MSVC is inherently challenging, efficient algorithms exist for specific graph classes. Gera et al. provide polynomial-time algorithms for MSVC for some subclasses of bipartite graphs \cite{gera2006results}. 
Rasmussen gave efficient algorithms for solving MSVC on caterpillar and split graphs \cite{rasmussen2006efficient}. 

In recent times, MSVC has attracted some attention in parameterized complexity. Aute and Panolan initiated the work on MSVC, and gave FPT algorithms with parameters vertex cover number, and size of clique modulator \cite{aute2025parameterized}.
The maximum cost of an edge in an optimal ordering is bounded by a function of vertex cover number. Liu et al. considered maximum cost of an edge as the parameter, and gave kernelization algorithm, and proved that MSVC is tractable with this parameter \cite{liu2025minimum}.
Zhang and Cao extended this result to hypergraphs; and proved that MSSC is tractable \cite{zhang2026minimum}.

\noindent
{\bf Organization of the paper.}
In the next section, we present our notations and some basic results on MSVC, which will be helpful in designing the algorithms for all kinds of graphs considered in this paper. 
We describe our FPT algorithms
in Section \ref{sec:cluster} and Section \ref{sec:clique} for parameterization by cluster modulator size +
maximum clique size and clique modulator size, respectively. The XP algorithm
is presented for WMSVC parameterized by clique modulator size in Section \ref{subsection:wtclique},
which in turn helps in developing an XP algorithm for MSVC parameterized by
split modulator size in Section \ref{sec:split}. We summarize our results and list some open
problems in Section \ref{sec:conclusion}. Proofs of results marked with \((\ast)\) appear in the appendix.

\section{Preliminaries and Notations}
Let $G$ be a graph with vertex set $V(G)$ and edge set $E(G)$. For $v\in V(G)$, let $G-v$ denote the subgraph induced by $V(G)\setminus\{v\}$. We write $N_G(u)$ for the open neighborhood of a vertex $u$ in $G$. For $M\subseteq V(G)$, let $N_M(u)=N_G(u)\cap M$.

Let $\phi$ be an ordering of $V(G)$ and let $X\subseteq V(G)$. The restriction of $\phi$ to $X$, denoted by $\left.\phi\right|_X$, preserves the relative order of the vertices in $X$ and renumbers their positions from $1$ to $|X|$. Formally, $\left.\phi\right|_X:X\to\{1,2,\ldots,|X|\}$ is the unique bijection such that, for all $u,v\in X$, $\phi(u)<\phi(v)$ if and only if $\left.\phi\right|_X(u)<\left.\phi\right|_X(v)$. We also write $u\prec_{\phi}v$ to mean $\phi(u)<\phi(v)$, omitting the subscript when the ordering is clear from the context. We say that an ordering $\phi$ of $V(G)$ is optimal if $\mu_G(\phi)$ is minimum over all orderings of $V(G)$.

\begin{definition}
    The right degree of a vertex $v$ in an ordering $\phi$ is defined as 
     $rd_{\phi}(v) = |\{u \in N_G(v): \phi(u) > \phi(v) \} |$ and the right modulator degree of  $v$ for a modulator $M$ in $\phi$ is defined as 
      $rm_{\phi}(v) = |\{u \in N_G(v) \cap M: \phi(u) > \phi(v) \} |.$ 
 \end{definition}

\begin{lemma}[\cite{aute2025parameterized}]
\label{rd}
   In an optimal ordering $\phi$ of MSVC, the sequence of right degrees of vertices is non-increasing.
\end{lemma}

\begin{figure}[h]
    \centering
    \includegraphics[width=0.65\linewidth]{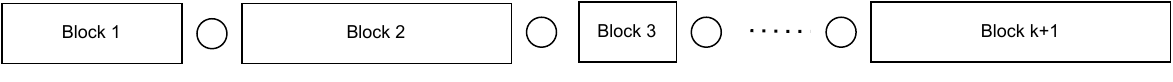}
    \caption{The $k+1$ blocks defined by a permutation of the $k$ modulator vertices.}
    \label{fig:blocks}
\end{figure}

\begin{lemma}[\cite{aute2025parameterized}]\label{l2}
    In an ordering $\phi$ of a graph $G$, locations of two consecutive non-adjacent vertices of equal right degrees can be swapped to get a new ordering with the same cost. 
\end{lemma}
Throughout the paper, $k$ denotes the modulator size. Our algorithms enumerate permutations of the modulator vertices. Each permutation defines $k+1$ blocks: one before the first modulator vertex, one between each pair of consecutive modulator vertices, and one after the last. The remaining vertices are assigned to these blocks, some of which may be empty. In \autoref{fig:blocks}, circles represent modulator vertices and rectangles represent the blocks.

An instance of Integer Quadratic Programming (IQP) consists of an integer matrix $Q\in\mathbb{Z}^{n\times n}$, an integer matrix $A\in\mathbb{Z}^{m\times n}$, and an integer vector $b\in\mathbb{Z}^{m}$. The objective is to find a vector $x\in(\mathbb{Z}^{+})^n$ that minimizes $x^{T}Qx$ subject to $Ax\leq b$. Let $L$ denote the total encoding length of the instance in bits.

Lokshtanov~\cite{lokshtanov2015parameterized} gave an algorithm to solve IQP in time $f(n,\alpha)L^{O(1)}$, where $\alpha$ is the largest absolute value of an entry of $Q$ and $A$, and $f$ is doubly exponential. Ari and Hildebrand~\cite{ari2026curvature} improve this dependence to $(n\alpha)^{O(n^2)}$.

\section{MSVC for Cluster Modulator Graphs}
\label{sec:cluster}

A \emph{cluster graph} is a disjoint union of cliques. A set $M\subseteq V(G)$ is a \emph{cluster modulator} of $G$ if $G-M$ is a cluster graph. Let $Q_1,\dots,Q_p$ be the cliques of $G-M$, let $k=|M|$, and let $d=\max_{j\in[p]}|V(Q_j)|$. 

Deleting a cluster modulator may leave arbitrarily many cliques. However, cliques with the same structure and neighborhoods in $M$ are interchangeable. We exploit this observation to group the cliques into a bounded number of types and obtain an algorithm parameterized by $k+d$.

Let $S_1,\dots,S_{2^k}$ enumerate all subsets of $M$. For a clique $Q$ of $G-M$, define its \emph{type vector} as $\tau(Q)=(a_1,\dots,a_{2^k})$, where $a_j$ is the number of vertices $v\in V(Q)$ satisfying $N_M(v)=S_j$. Thus, each coordinate records how many vertices of $Q$ have a particular neighborhood in $M$.

Two cliques are \emph{equivalent} if their type vectors are equal. Equivalently, $Q$ and $Q'$ are equivalent if there is a bijection $g:V(Q)\to V(Q')$ satisfying $N_M(v)=N_M(g(v))$ for every $v\in V(Q)$. This relation partitions the cliques into equivalence classes $A_1,\dots,A_t$.
The number of equivalence classes is bounded by $t\leq\binom{2^k+d}{d}$. Thus, it depends only on $k+d$, although each class may contain arbitrarily many cliques.

\paragraph{Parameter: Cluster Modulator Size + Maximum Clique Size}

We enumerate the $k!$ permutations of $M$. Fix one such permutation $\sigma_M=(v_1,\dots,v_k)$, and let $B_1,\dots,B_{k+1}$ be the resulting blocks, where $B_1$ precedes $v_1$ and $B_{k+1}$ follows $v_k$; see \autoref{fig:blocks}. For this permutation, we construct an integer quadratic program that determines how the clique vertices are distributed among the blocks.

Fix a labeling
of a representative clique from each class and transfer it to the other cliques using neighborhood-preserving bijections. If the cliques in $A_i$ have size $d_i$, a \emph{profile} for $A_i$ is a function $f:[d_i]\to[k+1]$ assigning each labeled vertex to a block. Let $\mathcal{F}_i$ denote the set of profiles for $A_i$. Then $|\mathcal{F}_i|=(k+1)^{d_i}\leq(k+1)^d$.

The following lemma allows us to count the cliques assigned to each profile without specifying their identities.

\begin{lemma}$(\ast)$
\label{lem:cluster-exchange}
Let $Q$ and $Q'$ be equivalent cliques, and let $g:V(Q)\to V(Q')$ be a neighborhood-preserving bijection. Given an ordering $\sigma$, let $\phi$ be obtained by exchanging the positions of $v$ and $g(v)$ for every $v\in V(Q)$. Then $\mu(\sigma)=\mu(\phi)$.
\end{lemma}

A profile determines the block of each vertex but does not specify the relative order of vertices within a block. We next establish an ordering property that makes their right degrees well defined.

\begin{lemma}$(\ast)$\label{lem:cluster-block-order}
Fix $\sigma_M$ and a profile for every clique. There exists a minimum-cost ordering consistent with these choices in which: the vertices of each clique within a block appear in non-increasing order of their right modulator degrees;
and all clique vertices within each block appear in non-increasing order of their right degrees. 
\end{lemma}

For each class $A_i$ and profile $f$, order its labeled 
vertices first by block and then, within each block, by non-increasing right modulator degree, breaking ties by label. Recall that, for each label $\ell$ there is a class $A_i$ such that one vertex per clique in $A_i$ is labeled $\ell$ and the labeling essentially describes the isomorphism between cliques in $A_i$. Let $r_{i,f}(\ell)$ denote the resulting right degree of labeled vertex $\ell$. This value counts the vertices of its clique following $\ell$ and its modulator neighbors following $B_{f(\ell)}$. It is therefore determined by $i$, $f$, and $\sigma_M$, and satisfies $0\leq r_{i,f}(\ell)\leq R$, where $R=k+d-1$.

We now describe the IQP. Let $x_{i,f}$ be a variable that encodes the number of cliques in $A_i$ assigned profile $f$.
These variables are nonnegative integers satisfying $\sum_{f\in\mathcal{F}_i}x_{i,f}=|A_i|$ for every $i\in[t]$.

For modulator vertex $v_\alpha$, with $\alpha\in[k]$, define the following constants that depend only on the permutation of the cluster modulator:
 $b_{i,f,\alpha}$ is the number of vertices in a clique of class $A_i$ following profile $f$ that are assigned to blocks $B_1,\dots,B_\alpha$; and $e_{i,f,\alpha}$ is the number of neighbors of $v_\alpha$ in that clique assigned to blocks $B_{\alpha+1},\dots,B_{k+1}$; and $rm_\alpha$ is 
 the right modulator degree of $v_\alpha$ in the fixed permutation of $M$.
We can now express the position $P_\alpha$ and the right degree $D_\alpha$ of $v_\alpha$ as follows:  
Using positions numbered from $1$, we obtain
\begin{equation*}
    P_\alpha=\alpha+\sum_{i=1}^{t}\sum_{f\in\mathcal{F}_i}
        b_{i,f,\alpha}\cdot x_{i,f}, \quad D_\alpha
    =rm_\alpha+\sum_{i=1}^{t}\sum_{f\in\mathcal{F}_i}
        e_{i,f,\alpha}\cdot x_{i,f}.
\end{equation*}
Edges whose earlier endpoint belongs to $M$ therefore contribute $\sum_{\alpha=1}^{k}P_\alpha \cdot  D_\alpha$.

To account for the remaining edges, we group clique vertices by their block and right degree. For profile $f$ of equivalence class $A_i$, block $j\in[k+1]$ and right degree $r\in\{0,\dots,R\}$, 
let $y_{i,f,j,r}$ 
be the number of vertices that receive some label $\ell$ for which $f(\ell)=j$ and $r_{i,f}(\ell)=r$.
The total number $h_{j,r}$ and starting location $q_{j,r}$ of such vertices are: 
\[
    h_{j,r}
    =\sum_{i=1}^{t}\sum_{f\in\mathcal{F}_i}
        y_{i,f,j,r}\cdot x_{i,f}, \quad \text{ and }\quad q_{j,r}
    =j+\sum_{b=1}^{j-1}\sum_{s=0}^{R}h_{b,s}
       +\sum_{s=r+1}^{R}h_{j,s}.
\]

Note that by \autoref{lem:cluster-block-order} these vertices occupy consecutive positions within $B_j$ and $q_{j,r}$ is the position of the first such vertex. 
The term $j$ accounts for the $j-1$ preceding modulator vertices and the use of one-based positions. The remaining terms count clique vertices in earlier blocks and vertices of larger right degree in the same block. If $h_{j,r}=0$, the value of $q_{j,r}$ does not affect the objective.
The contribution of this group  is
\(
    r\cdot \sum_{\ell=0}^{h_{j,r}-1}(q_{j,r}+\ell)
    =r\cdot \,h_{j,r}
       \left(q_{j,r}+\frac{h_{j,r}-1}{2}\right).
\)
Vertices of right degree zero contribute no cost, but their counts must be included when determining positions. We describe the IQP as follows:

\textbf{ Minimize } \(\displaystyle
    \sum_{\alpha=1}^{k} P_\alpha \cdot D_\alpha
    + \sum_{j=1}^{k+1}\sum_{r=0}^{R}
      r \cdot h_{j,r}
      \left(q_{j,r}+\frac{h_{j,r}-1}{2}\right)
\)
\medskip

{\bf Subject to}

\[
\begin{aligned}
    &\sum_{f\in\mathcal{F}_i}x_{i,f}=|A_i|
    &&\text{for all }i\in[t],\\[2pt]
    &P_\alpha
      =\alpha+\sum_{i=1}^{t}\sum_{f\in\mathcal{F}_i}
         b_{i,f,\alpha}\cdot x_{i,f}
    &&\text{for all }\alpha\in[k],\\[2pt]
    &D_\alpha
      =rm_\alpha+\sum_{i=1}^{t}\sum_{f\in\mathcal{F}_i}
         e_{i,f,\alpha}\cdot x_{i,f}
    &&\text{for all }\alpha\in[k],\\[2pt]
    &h_{j,r}
      =\sum_{i=1}^{t}\sum_{f\in\mathcal{F}_i}
         y_{i,f,j,r}\cdot x_{i,f}
    &&\text{for all }j\in[k+1],\ 0\leq r\leq R,\\[2pt]
    &q_{j,r}
      =j+\sum_{b=1}^{j-1}\sum_{s=0}^{R}h_{b,s}
         +\sum_{s=r+1}^{R}h_{j,s}
    &&\text{for all }j\in[k+1],\ 0\leq r\leq R.
\end{aligned}
\]
All variables are nonnegative integers. 

\begin{lemma}$(\ast)$\label{t3}
For a fixed permutation $\sigma_M$, the optimum value of the IQP equals the minimum MSVC cost among orderings whose restriction to $M$ is $\sigma_M$.
\end{lemma}

We bound the running time using the IQP algorithm of Ari and Hildebrand \cite{ari2026curvature}. 
Let $N=\sum_{i=1}^{t}|\mathcal{F}_i|$ be the number of profile-count variables, and define $B=\left[\binom{2^k+d}{d}-1\right](k+1)^d$. Then $N\leq B$.
By Corollary~1.2 of \cite{ari2026curvature}, for an IQP with $N$ variables, the running time for each permutation is,
$
    \bigl[B(k+d)d^2\bigr]^{O(B)}n^{O(1)}
    =\bigl[B(k+d)\bigr]^{O(B)}n^{O(1)}.
$

\begin{mainrestated}$(\ast)$
    Let $M$ be a cluster modulator of size $k$ such that every clique of $G-M$ has size at most $d$. Then
    MSVC is fixed-parameter tractable parameterized by $k+d$. 
\end{mainrestated}

\section{MSVC Parameterized by Clique Modulator}\label{sec:clique}
A set $M \subseteq V(G)$ is a \emph{clique modulator} of a graph $G$ if $G[V(G)\setminus M]$ is a clique. Let $M=\{v_1,v_2,\dots,v_k\}$ be a clique modulator, and let $Q=V(G)\setminus M$. Define an equivalence relation on $Q$ by declaring two vertices $u,w\in Q$ equivalent if $N_M(u)=N_M(w)$. This relation partitions $Q$ into equivalence classes $A_1,A_2,\dots,A_t$, with $t\leq 2^k$.

Aute and Panolan \cite{aute2025parameterized} showed that MSVC is fixed-parameter tractable when parameterized by the size of a clique modulator. Their algorithm enumerates the permutations of the modulator vertices. Each permutation defines $k+1$ blocks into which the vertices of $Q$ can be placed; see \autoref{fig:blocks}. Since vertices from a single equivalence class may be distributed across several blocks, their algorithm uses integer quadratic programming (IQP) to determine how many vertices from each class to assign to each block.
We present an FPT algorithm for the same parameterization that avoids IQP. The key ingredient is \autoref{thm:main}, which provides a structural property governing the distribution of vertices from each equivalence class among the blocks.

We first introduce the notation and ordering properties used in our algorithm. Fix a permutation $\sigma_M$ of $M$, and let $B_1,\dots,B_{k+1}$ denote the resulting blocks; see \autoref{fig:blocks}. For a vertex $u\in Q$ assigned to block $B_i$, define its \emph{right modulator degree}, $rm_{\sigma_M}(u,i)$, as the number of its neighbors in $M$ that occur after $B_i$. Vertices in the same equivalence class have identical neighborhoods in $M$ and therefore have the same right modulator degree within a given block. Accordingly, for an equivalence class $A$, we write $rm_{\sigma_M}(A,i)$ for this common value.

As shown in \cite{aute2025parameterized}, the vertices of an equivalence class assigned to a given block can be placed consecutively within that block without changing the cost of an optimal ordering. We call an optimal ordering \emph{nice} if, for every equivalence class $A$ and every block $B_i$, the vertices of $A$ assigned to $B_i$ appear consecutively. Henceforth, we restrict attention to nice optimal orderings.

Moreover, by Lemmas~5 and~6 of \cite{aute2025parameterized}, within each block, vertices from different equivalence classes appear in non-increasing order of their right modulator degrees.

\begin{theorem}\label{thm:main}
There exists a nice optimal ordering in which, for every equivalence class $A$ with $|A|>k$, all but at most $k$ vertices of $A$ lie in the first block containing vertices of $A$.
\end{theorem}

\begin{proof}
For an ordering $\phi$ and an equivalence class $A$ with $|A|>k$, let $A_1$ be the set of vertices of $A$ in the first block containing vertices of $A$. Define the \emph{deviation} of $A$ in $\phi$ as $D(A,\phi)=|A|-|A_1|$. For classes of size at most $k$, set $D(A,\phi)=0$. The deviation of $\phi$ is $D(\phi)=\sum_A D(A,\phi)$, where the sum ranges over all equivalence classes.

Choose a nice optimal ordering $\phi$ that minimizes $D(\phi)$. Suppose, for a contradiction, that some equivalence class $A$ satisfies $D(A,\phi)>k$. Let $B_i$ and $B_{i'}$, with $i<i'$, be the first two blocks containing vertices of $A$, and let $A_1$ and $A_2$ be the respective sets of vertices of $A$ in these blocks. Thus, $|A_1|<|A|-k$. Since $\phi$ is nice, the vertices of each of $A_1$ and $A_2$ appear consecutively within their respective blocks. Let $M'$ be the set of modulator vertices between $A_1$ and $A_2$. Since $i<i'$, we have $|M'|>0$.

\begin{claim}\label{claim:adj}
Every vertex of $M'$ is adjacent to every vertex of $A$.
\end{claim}

\begin{claimproof}
Suppose that some vertex $v\in M'$ is not adjacent to a vertex of $A$. Since all vertices of $A$ have the same neighborhood in $M$, the vertex $v$ has no neighbors in $A$.

Let $u$ be the first clique vertex appearing after $v$ in $\phi$, and let $h$ be the number of clique vertices appearing after $v$, including $u$. Since $Q$ is a clique, $rd_\phi(u)\geq h-1$. All vertices of $A\setminus A_1$ appear after $v$, so these $h$ clique vertices include at least $k+1$ vertices of $A$. The vertex $v$ is adjacent to none of these vertices and has at most $k-1$ neighbors in $M$. Hence,
\[
    rd_\phi(v)\leq h-(k+1)+(k-1)=h-2<rd_\phi(u).
\]
This contradicts \autoref{rd}, which states that right degrees are non-increasing along an optimal ordering, since $v\prec_\phi u$. \hfill $\diamond$
\end{claimproof}

Let $a_1$ be the last vertex of $A_1$, let $a_2$ be the first vertex of $A_2$, and write $\phi(a_1)=\ell$. Denote the vertices between $a_1$ and $a_2$, in their order of appearance, by $x_1,\dots,x_p$. Thus, $\phi(x_j)=\ell+j$ for $j\in[p]$, and $\phi(a_2)=\ell+p+1$. By the above Claim, every modulator vertex among the $x_j$ is adjacent to $A$. The remaining vertices belong to $Q\setminus A$ and are therefore also adjacent to $A$.

Write $rd_\phi(a_1)=d$ and $rd_\phi(x_j)=d_j$ for $j\in[p]$. Since $a_1$ and $a_2$ belong to the same equivalence class and every $x_j$ is adjacent to both, $rd_\phi(a_2)=d-p-1$.

Construct an ordering $\psi$ by moving $a_2$ immediately after $a_1$ and shifting each $x_j$ one position to the right. Then $\psi(a_2)=\ell+1$ and $\psi(x_j)=\ell+j+1$, while $rd_\psi(a_2)=d-1$ and $rd_\psi(x_j)=d_j-1$. All other vertices retain their positions and right degrees. Using the expression for the MSVC cost in terms of right degrees, we obtain
\begin{align*}
    \mu(\psi)-\mu(\phi)
    &=(\ell+1)(d-1)-(\ell+p+1)(d-p-1)\\
    &\quad+\sum_{j=1}^{p}
       \bigl[(\ell+j+1)(d_j-1)-(\ell+j)d_j\bigr]\\
    &=\sum_{j=1}^{p}d_j-pd+\frac{p(p+1)}{2}.
\end{align*}

Since $\phi$ is optimal, this difference is nonnegative. If it were zero, then $\psi$ would also be optimal. Moreover, moving $a_2$ into $B_i$ decreases the deviation of $A$ by one and leaves the deviations of all other classes unchanged. Thus, $D(\psi)=D(\phi)-1$. If necessary, we can rearrange vertices within each block to obtain a nice optimal ordering with the same block assignments and hence the same deviation. This contradicts the choice of $\phi$. Therefore,
\begin{equation}\label{eq:clique-shift-positive}
    \sum_{j=1}^{p}d_j-pd+\frac{p(p+1)}{2}>0.
\end{equation}

Now construct another ordering $\lambda$ from $\phi$ by moving $a_1$ immediately before $a_2$ and shifting each $x_j$ one position to the left. Then $\lambda(a_1)=\ell+p$ and $\lambda(x_j)=\ell+j-1$, while $rd_\lambda(a_1)=d-p$ and $rd_\lambda(x_j)=d_j+1$. All other vertices retain their positions and right degrees. Consequently,
\begin{align*}
    \mu(\lambda)-\mu(\phi)
    &=(\ell+p)(d-p)-\ell d\\
    &\quad+\sum_{j=1}^{p}
       \bigl[(\ell+j-1)(d_j+1)-(\ell+j)d_j\bigr]\\
    &=-\sum_{j=1}^{p}d_j+pd-\frac{p(p+1)}{2} <0,
\end{align*}
where the final inequality follows from \eqref{eq:clique-shift-positive}. This contradicts the optimality of $\phi$.

Hence, $D(A,\phi)\leq k$ for every equivalence class $A$ with $|A|>k$, proving the theorem.
\qed
\end{proof}

\begin{maintworestated}
    There is an FPT algorithm for MSVC  parameterized by the clique modulator size $k$, with running time $2^{O(k2^k)}n^{O(1)}$.
\end{maintworestated}

\begin{proof}
We enumerate all $k!$ permutations of the modulator vertices. Fix one such permutation, which defines $k+1$ blocks.

By \autoref{thm:main}, there exists a nice optimal ordering in which, for every large equivalence class $A$, at least $|A|-k$ vertices lie in the first block containing vertices of $A$. We therefore treat $|A|-k$ vertices of $A$ as a single group and enumerate its placement among the $k+1$ blocks. Since vertices in the same equivalence class are twins, only the number of vertices assigned to each block matters. The remaining $k$ vertices can be distributed among the $k+1$ blocks in $\binom{2k}{k}$ ways, by the stars-and-bars formula. Thus, we enumerate at most $(k+1)\binom{2k}{k}$ assignments for each large equivalence class.

For a small equivalence class $A$ of size $s\leq k$, we enumerate all distributions of its vertices among the $k+1$ blocks. There are $\binom{s+k}{k}\leq\binom{2k}{k}$ such distributions.

Since there are at most $2^k$ equivalence classes, the number of combined assignments for a fixed permutation of the modulator vertices is at most
\(
    \left[(k+1)\binom{2k}{k}\right]^{2^k}
\).
For each assignment, we construct an ordering by placing the vertices of each equivalence class consecutively within each block and arranging the classes in non-increasing order of their right modulator degrees, breaking ties arbitrarily. We compute the cost of each resulting ordering and return one of minimum cost.

By \autoref{thm:main} and the ordering properties established above, the enumeration includes an optimal ordering. Constructing and evaluating each candidate takes polynomial time in $n$. Accounting for the $k!$ permutations of the modulator vertices, the total running time is
\(
    k!\left[(k+1)\binom{2k}{k}\right]^{2^k}n^{O(1)}.
\)

Using $\binom{2k}{k}\leq 4^k$, our running-time bound simplifies to $2^{O(k2^k)}n^{O(1)}$. 
\qed
\end{proof}
Our algorithm improves the running-time bound from
$2^{O(k^3 4^k)}n^{O(1)}$ to $2^{O(k2^k)}n^{O(1)}$.

\section{MSVC Parameterized by Split Modulator Size}\label{sec:split}
A graph $G=(V,E)$ is a \emph{split graph} if its vertex set can be partitioned
into a clique $Q$ and an independent set $I$. 
MSVC can be solved in polynomial time on split graphs
\cite{rasmussen2006efficient} in the following way. Repeatedly select a clique vertex of maximum
degree in the remaining graph, place it next in the ordering, and delete it,
breaking ties arbitrarily, and append
the vertices of $I$ in any order. In the resulting optimal ordering, every
independent set vertex has right degree zero.

A set $M\subseteq V$ is a \emph{split modulator} if $G-M$ is a split graph.
Given $M$, fix a partition $V=Q\cup I\cup M$, where $Q$ induces a clique and
$I$ is independent. 
Let $|Q|=n_1$, $|I|=n_2$, $|M|=k$, and $|V|=n$. For these
graphs, the greedy algorithm above need not be optimal, and independent
vertices may have positive right degree in an optimal ordering. We develop an
XP-time algorithm parameterized by $k$.

We briefly outline our approach. We divide the given instance into two parts and prove that the first contains at most \(O(k^3)\) vertices from the independent set, while the second contains only \(2k\) vertices from the clique. We then solve the two resulting instances separately and combine their solutions to obtain an optimal ordering of the original graph.

Given an ordering $\phi$, a \emph{group} is a maximal consecutive sequence of vertices belonging to the same
part of $Q$, $I$, or $M$. A group may contain a single vertex.
For each $v\in V$, define
$rc_{\phi}(v)=|\{u\in N(v)\cap Q:\phi(u)>\phi(v)\}|,$ and $
q_{\phi}(v)=|\{u\in Q:\phi(u)>\phi(v)\}|.$
These quantities are, respectively, the
right clique degree, and the number of clique vertices to
the right of $v$. The \emph{surplus} of $v$ is $s_{\phi}(v)=rd_{\phi}(v)-q_{\phi}(v)$. For $u\in I$, all neighbors belong to $Q\cup M$, so $rd_{\phi}(u)\le q_{\phi}(u)+k$ and $s_{\phi}(u)\le k$. More precisely,
\begin{equation}
\label{eq:clique-nonneighbors}
q_{\phi}(u)-rc_{\phi}(u)
=rd_{\phi}(u)-rc_{\phi}(u)-s_{\phi}(u)
\le k-s_{\phi}(u).
\end{equation}
Thus, an independent set vertex of surplus $s$ has at most
$k-s$ clique non-neighbors to its right.

To resolve ties between optimal orderings, define the \emph{potential} as $P(\phi)=\sum_{u\in
I}\phi(u)$. The structural arguments use optimal orderings maximizing $P$. We
omit the subscript $\phi$ when the ordering is clear. We also use two
properties of optimal MSVC orderings: right degrees are non-increasing, and
vertices with equal right degree are pairwise nonadjacent
\cite{aute2025parameterized}. Consequently, each equal-degree block can be
permuted without changing the cost or the right degrees. In particular, clique
vertices have strictly decreasing right degrees.

We begin with two exchange tools and an observation about restricting an
optimal ordering to its remaining vertices.

\begin{lemma}$(\ast)$
\label{lem:unit-degree-exchange}
Let $\phi$ be an optimal ordering, and let $u,v\in V$
satisfy $\phi(u)<\phi(v)$, $uv\in E$, and
$rd_{\phi}(u)=rd_{\phi}(v)+1$. Neither endpoint has a
neighbor strictly between them, and swapping $u$ and $v$
preserves the cost. 
\end{lemma}

\begin{lemma}$(\ast)$
\label{lem:insertion-estimate}
Let $\phi$ be an optimal ordering. Let $x\in I$ and
$y\in Q\cup M$ satisfy $\phi(x)<\phi(y)$, and set $\ell=\phi(y)-\phi(x)$. If $y$ has at least $r$ neighbors at positions
$\phi(x),\ldots,\phi(y)-1$, then $\frac{r(r+1)}2 \le \ell\bigl(rd_{\phi}(x)-rd_{\phi}(y)\bigr)$. 
\end{lemma}

\begin{lemma}$(\ast)$
\label{lem:g-v}
Let $\phi$ be an optimal ordering of $G$, and let $v$
be its first vertex. The ordering $\psi$ of $G-v$ defined
by $\psi(x)=\phi(x)-1$ is optimal for $G-v$. 
\end{lemma}

We next establish the surplus properties of optimal orderings maximizing $P$.

\begin{lemma}$(\ast)$
\label{lem:zero-degree-placement}
In an optimal ordering $\phi$ maximizing $P(\phi)$,
every independent set vertex of right degree zero occurs
after every vertex in $Q\cup M$. 
\end{lemma}

\begin{lemma}$(\ast)$
\label{lem:surplus-monotonicity}
Let $\phi$ be an optimal ordering maximizing $P(\phi)$. The surplus is non-increasing along the independent vertices. That is, for any $u,v\in I$ with $\phi(u)<\phi(v)$, we have $s_{\phi}(u)\ge s_{\phi}(v)$. 
\end{lemma}

The next lemma isolates the degree drop at the first right clique neighbor of
an independent set vertex. It will be used both for zero surplus and for strict
surplus decrease.

\begin{lemma}
\label{lem:first-clique-neighbor-drop}
Let $\phi$ be an optimal ordering maximizing $P(\phi)$.
Let $a\in I$, and suppose that $a$ has a clique neighbor
after it. Let $x\in Q$ be the first such neighbor, and
let $t$ be the number of clique vertices strictly between
$a$ and $x$. Then $rd(x)\le rd(a)-t-2$ and $s(x)\le s(a)-1$.
\end{lemma}

\begin{proof}
List the clique vertices strictly between $a$ and $x$
as $x_1,\ldots,x_t$. Suppose, for a contradiction, that $rd(x)\ge rd(a)-t-1$.
If $t\ge1$, the strict decrease of right degrees along the clique vertices
gives $rd(x_1)\ge rd(a)-1$. On the other hand,
$rd(x_1)\le rd(a)$ by monotonicity. Equality would allow
us to swap $a$ and $x_1$ in their equal-degree block,
preserving the cost and increasing $P$. Hence
$rd(x_1)=rd(a)-1$. The strict decreases of right degrees along
$x_1,\ldots,x_t,x$, together with the assumed bound on
$rd(x)$, now force $rd(x_j)=rd(a)-j$ for $1\le j\le t$ and $rd(x)=rd(a)-t-1$.
Apply \autoref{lem:unit-degree-exchange} to swap $x$
successively with $x_t,x_{t-1},\ldots,x_1$. Each swap
preserves the cost and $P$, since both vertices belong
to $Q$, and increases the right degree of $x$ by one.
At each step, the next clique vertex to be exchanged
precedes both swapped positions, so its right degree
is unchanged. Thus the exchange lemma applies at every
step. After the swaps, $x$ occupies the original position
of $x_1$ and has right degree $rd(a)-1$. The right degree
of $a$ is unchanged.

If $t=0$, the same equality holds without any preliminary swaps: adjacency and
monotonicity give $rd(x)\le rd(a)-1$, and the assumed bound gives equality.

In either case, $a$ precedes its neighbor $x$ and their
right degrees differ by one. By
\autoref{lem:unit-degree-exchange}, swapping them preserves
the cost. This moves $a\in I$ to a later position without
changing the positions of any other independent vertices,
contradicting maximality of $P$. Therefore, $rd(x)\le rd(a)-t-2$.
Finally, $q(a)=q(x)+t+1$, so
$s(x)=rd(x)-q(x)\le rd(a)-t-2-\bigl(q(a)-t-1\bigr)=s(a)-1$.
\qed 
\end{proof}

\begin{lemma}$(\ast)$
\label{lem:zero-surplus}
Let $\phi$ be an optimal ordering maximizing $P(\phi)$.
Every $u\in I$ has $s(u)\ge0$. If $s(u)=0$, then $rc(u)=0$ and $q(u)\le k$. In particular, $q(u)>k$ implies $s(u)>0$. 
\end{lemma}

\begin{lemma}
\label{lem:surplus}
Let $\phi$ be an optimal ordering maximizing $P(\phi)$.
Let $a\in I$ be the last vertex of an independent group,
with $s(a)>0$. Suppose that the first clique neighbor
$x\in Q$ after $a$ and the first independent set vertex $b\in I$ after $x$, both exist. Then $s(b)<s(a)$.
\end{lemma}

\begin{proof}
By \autoref{lem:first-clique-neighbor-drop}, $s(x)\le s(a)-1$. It remains to show that $s(b)\le s(x)$. Let $y_1,\ldots,y_p$ be the clique vertices strictly
between $x$ and $b$. Then $q(b)=q(x)-p$.

If $p\ge1$, strict decrease along the clique vertices and right-degree
monotonicity give $rd(b)\le rd(y_p)\le rd(x)-p$. The same inequality holds for
$p=0$ by monotonicity. Consequently, $s(b)=rd(b)-q(x)+p \le
rd(x)-q(x)=s(x)<s(a)$.
\qed
\end{proof}

We will also need a bound on the number of clique vertices between independent vertices
of the same positive surplus.

\begin{lemma}$(\ast)$
\label{lem:equal-surplus-clique-bound}
Let $\phi$ be an optimal ordering maximizing $P(\phi)$.
Let $u,z\in I$ satisfy $\phi(u)\le\phi(z)$ and $s(u)=s(z)=s>0$. Then at most $k-s$ clique vertices occur strictly between $u$ and $z$.
\end{lemma}

We now divide the ordering at a threshold $T$. Assume $k\ge1$.
If $|Q|\ge2k$, let $T$ be the position of the $(|Q|-2k+1)$st
clique vertex; otherwise, set $T=0$. We use \emph{before threshold $T$}
for positions at most $T$, and \emph{after threshold $T$} for positions
greater than $T$.

\begin{figure}[t]
  \centering
  \begin{tikzpicture}[
  x=1cm, y=1cm,
  line width=0.55pt,
  every node/.style={font=\small},
  band/.style={rounded corners=2pt, line width=0.6pt},
  vertex/.style={circle, draw=black, fill=white,
    minimum size=4pt, inner sep=0pt, outer sep=0pt},
  brace/.style={decorate,
    decoration={brace, mirror, amplitude=3pt}, line width=0.65pt}
]
  \draw[band, draw=green!55!black, fill=green!15]
    (0,-0.15) rectangle (6.20,0.15);
  \draw[band, draw=violet!60!gray, fill=violet!17]
    (6.45,-0.15) rectangle (12,0.15);
  \draw[band, draw=red!65!black, fill=red!20]
    (6.45,-0.15) rectangle (8.10,0.15);

  \draw[blue!75!violet, line width=1.2pt]
    (6.20,-0.35) -- (6.20,0.40);

  \draw (3.50,0)
    .. controls (3.50,0.65) and (7.05,0.65) .. (7.05,0);
  \draw (3.50,0)
    .. controls (3.50,1.00) and (11.35,1.00) .. (11.35,0);

  \node[vertex] at (3.50,0) {};
  \node[vertex] at (7.05,0) {};
  \node[vertex] at (11.35,0) {};

  \draw[brace] (0,-0.27) -- (6.15,-0.27)
    node[midway, below=4pt] {$V_1$};
  \draw[brace] (6.45,-0.27) -- (12,-0.27)
    node[midway, below=4pt] {$V_2$};
  \node[anchor=north, inner sep=1pt] at (6.20,-0.39) {$T$};
\end{tikzpicture}
  \caption{Threshold $T$ creating weighted Graph instance $G_1$, and Graph $G_2$.}
  \label{fig:threshold}
\end{figure}
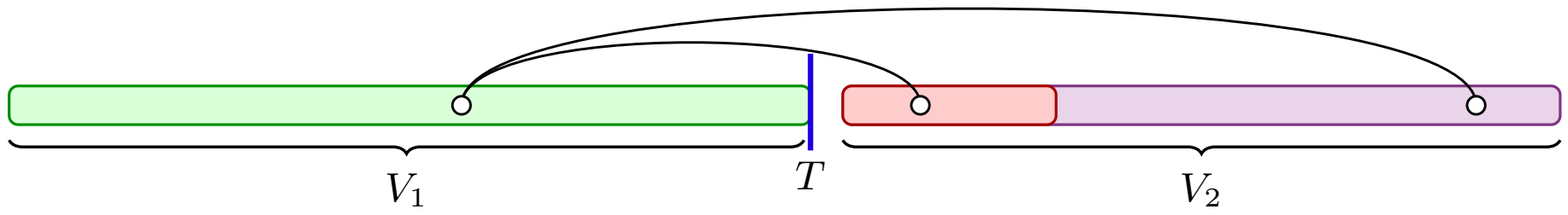
\autoref{fig:threshold} represents the division of the ordering. The edges across $V_1$ and $V_2$ show the requirement of the weighted graph $G_1$. Even though there can be a large number of vertices in $V_2$, its vertex cover is small and is shown in red. We first bound the vertices of positive right degree after $T$.

\begin{lemma}$(\ast)$
\label{lem:after-threshold-bound}
Let $k\ge1$, let $\phi$ be an optimal ordering maximizing
$P(\phi)$, and let $T$ be defined as above.
The number of vertices of positive right degree after
threshold $T$ is $O(k\log(k+1))$.
In particular, the number of independent vertices of
positive surplus after $T$ is $O(k\log(k+1))$. 
\end{lemma}

To bound the independent vertices before $T$, we apply the insertion estimate to a later
clique vertex with many neighbors in a largest set of independent vertices of
equal surplus.

\begin{lemma}
\label{lem:positive-surplus-bound}
There exists an optimal ordering maximizing $P(\phi)$, in which the number
of independent vertices with positive surplus is $O(k^3)$.
\end{lemma}

\begin{proof}
Choose an optimal ordering $\phi$ maximizing $P(\phi)$, and let $T$ be the
threshold defined above. By \autoref{lem:after-threshold-bound}, the number of
independent vertices of positive right degree after $T$ is at most
$\beta=O(k\log(k+1))$. For each $s\in\{1,\ldots,k\}$, let $A_s=\{u\in I:\phi(u)\le
T,\ s(u)=s\}$. Every independent set vertex before $T$ has at least $2k$ clique vertices to its right and therefore has positive surplus ($\leq k$) by \autoref{lem:zero-surplus}.  Thus the sets
$A_1,\ldots,A_k$ partition the independent vertices before $T$.
If all these sets are empty, the bound after $T$ proves the lemma. Otherwise,
choose a largest set $A_s$ and write $\alpha=|A_s|$. Let $a,z\in I$ be its first
and last vertices, respectively.

Since $z$ occurs before $T$, at least $2k$ clique vertices occur after $z$.
Let $C\subseteq Q$ consist of the first $2k$ such vertices.
By Equation \eqref{eq:clique-nonneighbors}, every $u\in A_s$ has at most $k-s$ clique
non-neighbors to its right. All vertices of $C$ occur after $u$, so $u$ has at
least $2k-(k-s)=k+s\ge k$ neighbors in $C$. There are therefore at least $\alpha k$
edges between $A_s$ and $C$. Since $|C|=2k$, some clique vertex $v\in C$ has
at least $\alpha/2$ neighbors in $A_s$.

By \autoref{lem:equal-surplus-clique-bound}, at most $k-s$ clique vertices occur between $a$ and $z$. By the choice of $C$, at most $2k$ clique vertices
occur after $z$ and at or before $v$. Thus, $q(v)\ge q(a)-(3k-s)$. Because $v\in Q$, it is adjacent to every clique vertex
after it, giving $rd(v)\ge q(v)$. Since $rd(a)=q(a)+s$,
we obtain
\begin{equation}
\label{eq:positive-surplus-degree-gap}
rd(a)-rd(v)
\le q(a)+s-\bigl(q(a)-(3k-s)\bigr)
=3k.
\end{equation}

Let $D=\phi(v)-\phi(a)$. We now bound the number of vertices at positions
$\phi(a),\ldots,\phi(v)-1$. There are at most $k\alpha$ independent vertices from
before $T$, because each $A_t$ has size at most $\alpha$. By
\autoref{lem:zero-degree-placement}, every independent set vertex in this interval
has positive right degree, since it precedes $v\in Q$. Hence, at most $\beta$ of
them occur after $T$.
The interval also contains at most $3k-s$ clique vertices and at most $k$
modulator vertices. Consequently, $D\le k\alpha+\beta+4k$.

Every vertex of $A_s$ is in a position from $\phi(a)$
through to $\phi(v)-1$, and
at least $\alpha/2$ of them are adjacent to $v$.
Applying \autoref{lem:insertion-estimate} with
$x=a\in I$, $y=v\in Q$, and $r=|N(v)\cap A_s|\ge \alpha/2$, and then
using \eqref{eq:positive-surplus-degree-gap} and the bound
$D\le k\alpha+\beta+4k$, 
yields
\[
\frac{\alpha^2}{8}
\le \frac{r(r+1)}2
\le D\bigl(rd(a)-rd(v)\bigr)
\le 3k(k\alpha+\beta+4k).
\]
Therefore, $\alpha^2 \le24k^2\alpha+24k\beta+96k^2$.
If $\alpha\le48k^2$, then $\alpha=O(k^2)$ immediately.
Otherwise, $24k^2\alpha<\alpha^2/2$, and hence $\frac{\alpha^2}{2} \le24k\beta+96k^2 =O(k^2\log(k+1))$, which gives $\alpha=O(k\sqrt{\log(k+1)})=O(k^2)$.

Thus, the number of independent vertices before $T$ is at most $k\alpha=O(k^3)$. By
\autoref{lem:after-threshold-bound}, at most $O(k\log(k+1))$ independent vertices
of positive surplus occur after $T$. The total is therefore $O(k^3)$.
\qed
\end{proof}

We now turn to the algorithm. The structural bound limits the independent
vertices before $T$, while the definition of $T$ limits the clique vertices
after it. We first explain how a fixed partition at $T$ separates the
optimization.

Fix a partition $V=V_1\cup V_2$, where $V_1$ consists of the vertices at
positions at most $T$ in an optimal ordering, and $V_2$ consists of the
vertices after $T$. Thus, $|V_1|=T$. We construct an optimal ordering of $G$
by solving two instances corresponding to this partition.

Let $G_2=G[V_2]$. By repeated application of \autoref{lem:g-v}, the ordering
after $T$, with its positions shifted down by $T$, is optimal for $G_2$. We
may therefore replace it with any optimal ordering of $G_2$.

Let $\alpha=\alpha(k)=O(k^3)$ be the bound from
\autoref{lem:positive-surplus-bound}. We enumerate sets
$I_1\subseteq I$, $Q_2\subseteq Q$, and $M_1\subseteq M$ with
$|I_1|\le\alpha$ and $|Q_2|\le2k$, defining
\[
V_1=(Q\setminus Q_2)\cup I_1\cup M_1,
\qquad V_2=V(G)\setminus V_1.
\]
Indeed, the threshold bounds guarantee that an optimal ordering induces
a partition with $|I_1|\le\alpha$ and $|Q_2|\le2k$, and our enumeration
includes every such choice of $I_1$, $Q_2$, and $M_1$. 
For each candidate partition, we set $T=|V_1|$ and solve the following two instances.

For $G_1=G[V_1]$, assign each vertex the weight
$w(v)=|N(v)\cap V_2|$ and define the cost 
\[
\overline\mu_{G_1}(\rho)=\sum_{v\in V_1}\rho(v)
\bigl(rd_\rho(v)+w(v)\bigr),
\]
where $rd_\rho(v)$ counts right neighbors within $V_1$.
These weights account for edges between the two parts.
Since $I_1\cup M_1$ is a clique modulator of size at most
$\alpha+k=O(k^3)$, we solve this weighted instance using the XP algorithm for WMSVC parameterized by the clique modulator size.
For $G_2=G[V_2]$, the set $Q_2\cup(M\setminus M_1)$ is a vertex
cover of size at most $3k$. For $k\ge1$, an optimal ordering therefore
has at most $\beta=\lceil6k\log_2(3k)\rceil$ vertices of positive
right degree~\cite[Theorem~1.1]{zhang2026minimum}.
We find an optimum by enumerating ordered subsets of at most $\beta$
vertices whose complements are independent, appending each complement
in arbitrary order. This takes $n^{O(k\log(k+1))}$ time.

Concatenating optimal orderings $\sigma_1$ and $\sigma_2$ of the two
instances gives
\begin{equation}
\label{eq:split-decomposition-cost}
\overline\mu_G(\sigma_1\cdot\sigma_2)
=\overline\mu_{G_1}(\sigma_1)+\mu_{G_2}(\sigma_2)+T|E(G_2)|.
\end{equation}
The last term is constant for a fixed partition. We return the
minimum-cost concatenation over all enumerated partitions.

\begin{theorem}
\label{thm:split-modulator-xp}
MSVC admits an XP-time algorithm parameterized by the
size of a given split modulator.
\end{theorem}
\begin{proof}[Proof sketch]
For $k\ge1$, an enumerated partition matches an optimal ordering;
Equation \eqref{eq:split-decomposition-cost} then ensures that solving its two
instances independently preserve optimality.
There are $2^k n^{O(k^3)}$ partitions, and both subproblems take XP
time parameterized by $k$. Thus, the total running time is
$f(k)n^{g(k)}$ for computable functions $f,g$.
\qed
\end{proof}

\section{Conclusion}\label{sec:conclusion}
In this work, we study MSVC on different types of graphs that admit a small modulator whose removal leaves a large clique, a large number of small cliques, or a split graph. We give FPT algorithms for MSVC, parameterized by the cluster modulator size and the maximum clique size. An interesting open question is whether MSVC is FPT with only a single parameter: the cluster modulator size. Parameterized by the clique modulator size, we present an FPT algorithm for MSVC on unweighted graphs and an XP algorithm on vertex-weighted graphs. With an algorithm for WMSVC as a key ingredient, we further obtain an XP algorithm parameterized by the split modulator size. Determining whether MSVC is FPT parameterized by the split modulator size is left as an open question. 
The time complexities for MSVC and WMSVC, parameterized by the clique modulator size, suggest that vertex-weighted versions, in general, may offer sharp boundaries between tractability and hardness.

%
%
%
%


\bibliographystyle{splncs04}
\bibliography{ref.bib}
\appendix
\newpage

\section{WMSVC parameterized by clique modulator size}
\label{subsection:wtclique}

The preceding algorithm is effective when the clique modulator is small. However, even graphs with a simple structure can have a large clique modulator. Consider a large clique with many pendant vertices attached to it. The pendant vertices can make the clique modulator arbitrarily large, while the clique itself forces the vertex cover to be large. Thus, neither parameter fully captures the simplicity of this structure. This motivates a weighted formulation that represents pendant vertices by weights on their neighbors, allowing us to exploit the structure of the remaining graph.

To see why this formulation is natural, consider a connected graph with at least three vertices. There exists an optimal ordering in which every pendant vertex appears after its unique neighbor. Indeed, suppose that a pendant vertex $u$ appears before its neighbor $v$. Swapping $u$ and $v$ leaves the contribution of the edge $uv$ unchanged and cannot increase the contribution of any other edge: every other edge incident to $v$ is covered no later than before, while all remaining edges are unaffected. Applying this exchange whenever necessary yields an optimal ordering in which every pendant vertex has right degree zero.

We can then move all pendant vertices to the end of the ordering without increasing the cost. Although these vertices have right degree zero, their incident edges still contribute to the objective: each pendant edge contributes the position of its non-pendant endpoint. We can therefore delete all pendant vertices and assign each remaining vertex $x$ a weight $w(x)$ equal to the number of deleted pendant neighbors of $x$.

Let $G'$ be the resulting vertex-weighted graph. For an ordering $\phi$ of $V(G')$, define the cost
\[
    \overline{\mu}_{G'}(\phi)
    =\sum_{x\in V(G')}\phi(x)\bigl[rd_\phi(x)+w(x)\bigr].
\]
The term $\phi(x)rd_\phi(x)$ accounts for edges retained in $G'$, while $\phi(x)w(x)$ accounts for the deleted pendant edges incident to $x$. Every ordering of $G'$ extends to an ordering of the original graph with the same cost by appending the deleted vertices. Conversely, the original graph has an optimal ordering of this form. Hence, the weighted instance has the same optimal value as the original instance. In particular, a clique with arbitrarily many pendant vertices reduces to a weighted clique.

Motivated by this reduction, we study MSVC on vertex-weighted graphs, called WMSVC, with a small clique modulator. Let $G$ have nonnegative vertex weights, let $M$ be a clique modulator of size $k$, and let $Q=V(G)\setminus M$. As before, we enumerate the $k!$ permutations of $M$. Each permutation defines $k+1$ blocks into which the vertices of $Q$ must be placed.

We partition $Q$ into at most $2^k$ equivalence classes according to neighborhoods in $M$: two vertices $u,v\in Q$ belong to the same class if $N_M(u)=N_M(v)$. For each permutation of $M$, we enumerate the number of vertices from each equivalence class assigned to each block.

Unlike in the unweighted setting, these counts alone do not determine the cost. Vertices in the same equivalence class are twins in the underlying graph, so swapping them leaves the edge contribution unchanged. Their weights, however, may differ, and swapping them can change the weighted contribution. We must therefore determine which vertices of a class occupy its assigned positions, rather than choose them arbitrarily. The following lemma provides the ordering property needed to make this assignment.

\begin{lemma}\label{lem:weight}
In every optimal ordering of WMSVC, on a vertex-weighted graph, with clique modulator size as parameter, the vertices of each equivalence class appear in non-increasing order of their weights. Vertices of equal weight within a class may be exchanged without changing the cost.
\end{lemma}

\begin{proof}
Let $u_1,u_2$ belong to the same equivalence class, with $\phi(u_1)=p<q=\phi(u_2)$ in an optimal ordering $\phi$. Suppose that $w(u_1)<w(u_2)$, and let $\psi$ be obtained by swapping $u_1$ and $u_2$. Since these vertices are twins in the underlying graph, the total edge contribution remains unchanged. Hence,
\begin{align*}
    \overline\mu(\psi)-\overline\mu(\phi)
    &=p\,w(u_2)+q\,w(u_1)-p\,w(u_1)-q\,w(u_2)\\
    &=(p-q)\bigl[w(u_2)-w(u_1)\bigr]<0,
\end{align*}
contradicting the optimality of $\phi$. If $w(u_1)=w(u_2)$, the same calculation gives $\overline\mu(\psi)=\overline\mu(\phi)$.
\qed
\end{proof}

By \autoref{lem:weight}, once the number of vertices from an equivalence class assigned to each block is fixed, we can determine the assignment by sorting the class in non-increasing order of weight. If $a_j$ vertices are assigned to block $B_j$, we assign the first $a_1$ vertices to $B_1$, the next $a_2$ vertices to $B_2$, and so on, breaking ties arbitrarily.

It remains to determine the order of vertices within each block. We first record the weighted analogue of \autoref{rd}.

\begin{lemma}
In every optimal ordering $\phi$ of WMSVC on a vertex-weighted graph, the values $rd_\phi(u)+w(u)$ are non-increasing along the ordering.
\end{lemma}

\begin{proof}
Let $u$ and $v$ occupy consecutive positions in $\phi$, with $u$ preceding $v$, and let $\psi$ be obtained by swapping them. Write $\varepsilon=1$ if $uv\in E(G)$ and $\varepsilon=0$ otherwise. Then
\[
    \overline\mu(\psi)-\overline\mu(\phi)
    =\bigl[rd_\phi(u)+w(u)\bigr]
     -\bigl[rd_\phi(v)+w(v)\bigr]-\varepsilon.
\]
If $rd_\phi(u)+w(u)<rd_\phi(v)+w(v)$, this difference is negative, contradicting optimality.
\qed
\end{proof}

In the unweighted setting, vertices within a block are ordered by their right modulator degrees, and vertices from the same equivalence class can be placed consecutively. With vertex weights, the appropriate sorting key is the sum of the right modulator degree and the weight. Consequently, vertices from different equivalence classes may interleave.

\begin{lemma}\label{lem:blocksort}
Fix a permutation $\sigma_M$ of the modulator vertices and an assignment of clique vertices to the resulting blocks. Among the orderings consistent with this assignment, the cost is minimized by sorting the vertices in each block $B_j$ in non-increasing order of $rm_{\sigma_M}(u,j)+w(u)$, breaking ties arbitrarily.
\end{lemma}

\begin{proof}
For a vertex $u$ assigned to $B_j$, write $s_j(u)=rm_{\sigma_M}(u,j)+w(u)$. Suppose that $u_1,u_2\in B_j$ occupy positions $p<q$, respectively, and let $\psi$ be obtained from $\phi$ by swapping them.

The total contribution of edges within the clique is unchanged, since the set of positions occupied by clique vertices remains the same. Edges from $u_1$ or $u_2$ to modulator vertices preceding $B_j$ also retain their contributions. Only the vertex-weight terms and edges to modulator vertices following $B_j$ can change. Therefore,
\begin{align*}
    \overline\mu(\psi)-\overline\mu(\phi)
    &=p\,s_j(u_2)+q\,s_j(u_1)
      -p\,s_j(u_1)-q\,s_j(u_2)\\
    &=(p-q)\bigl[s_j(u_2)-s_j(u_1)\bigr].
\end{align*}
If $s_j(u_1)<s_j(u_2)$, swapping the vertices strictly decreases the cost. If their sorting keys are equal, the cost is unchanged. Repeatedly exchanging adjacent inverted pairs therefore produces the required order without increasing the cost.
\qed
\end{proof}

Together, \autoref{lem:weight} and \autoref{lem:blocksort} reduce the search to the number of vertices from each equivalence class assigned to each block. The \autoref{lem:weight} determines which vertices occupy the assigned blocks, and \autoref{lem:blocksort} determines their order within each block.

Let $A_1,\dots,A_t$ be the equivalence classes, where $t\leq 2^k$, and let $n_1=|Q|$. For each permutation of $M$, we enumerate all nonnegative counts $a_{i,j}$ satisfying $\sum_{j=1}^{k+1}a_{i,j}=|A_i|$ for every $i\in[t]$. Class $A_i$ has $\binom{|A_i|+k}{k}$ possible distributions. Thus, the total number of combined assignments for a fixed permutation is
\[
    \prod_{i=1}^{t}\binom{|A_i|+k}{k}
    \leq \binom{n_1+k}{k}^{t}
    \leq \binom{n_1+k}{k}^{2^k}.
\]

For each combined assignment, we distribute the vertices of each class among the blocks in non-increasing order of weight, as prescribed by \autoref{lem:weight}. We then sort the vertices within each block according to \autoref{lem:blocksort}. This constructs a complete ordering, whose weighted cost can be evaluated in polynomial time. We repeat this procedure for all $k!$ permutations of $M$ and return an ordering of minimum cost.

To establish correctness, consider an optimal ordering. Our enumeration includes its permutation of $M$ and its counts for every class and block. By \autoref{lem:weight}, the resulting vertex assignments agree with those of an optimal ordering, up to exchanges of equal-weight vertices within a class. By \autoref{lem:blocksort}, sorting each block cannot increase the cost. Hence, the algorithm constructs an optimal ordering.

The total running time is
$   k!\binom{n_1+k}{k}^{2^k}n^{O(1)}.$
Since the exponent of $n$ depends only on $k$, this yields an XP algorithm parameterized by the clique modulator size.

\section{Proofs from Section~\ref{sec:cluster}}
\subsection{Proof of \autoref{lem:cluster-exchange}}
\begin{proof}
The map exchanging each $v$ with $g(v)$ and fixing all other vertices is an automorphism of $G$. Corresponding vertices have identical neighborhoods in $M$, and neither clique has neighbors in any other clique of $G-M$. 
The exchange thus preserves the cost.
\qed
\end{proof}
\subsection{Proof of \autoref{lem:cluster-block-order}}
\begingroup
\def\thelemma{\ref{lem:cluster-block-order}}
\begin{lemma}
Fix $\sigma_M$ and a profile for every clique. There exists a minimum-cost ordering consistent with these choices in which: 
\begin{itemize}
    \item the vertices of each clique within a block appear in non-increasing order of their right modulator degrees;
    \item all clique vertices within each block appear in non-increasing order of their right degrees.
    \end{itemize}
\end{lemma}
\addtocounter{lemma}{-1}
\endgroup

\begin{proof}
Consider two vertices $u$ and $v$ from the same clique and block, occupying positions $a<b$. Let $rm(u)$ and $rm(v)$ be their right modulator degrees. Swapping these vertices leaves the total contribution of edges within their clique unchanged, since the set of positions occupied by that clique remains the same. Contributions from edges to modulator vertices preceding the block are also unchanged. Hence, the change in cost is $(a-b)(rm(v)-rm(u))$.

If $rm(u)<rm(v)$, the exchange decreases the cost; if the values are equal, it preserves the cost. Repeated exchanges, therefore, establish the first property, with ties broken by the fixed labels.

Once these relative orders are fixed, every clique vertex has a fixed right degree: its right neighbors are the later vertices of its clique and its modulator neighbors following the block. Within a clique, right degrees strictly decrease along its vertices in a fixed block, since the number of later clique vertices decreases while the right modulator degree does not increase.

We can therefore merge the clique orders within each block into a sequence of non-increasing right degrees while preserving the relative order inside every clique. This leaves all right degrees unchanged, as well as the positions of the modulator vertices. Assigning larger right degrees to earlier positions minimizes the remaining cost, proving the second property.
\qed
\end{proof}
\subsection{Proof of \autoref{t3}}
\begin{proof}
Consider a feasible IQP solution. For every class $A_i$, assign exactly $x_{i,f}$ cliques to profile $f$. The identities of the chosen cliques do not matter, by \autoref{lem:cluster-exchange}. Order each clique as specified above, and merge the clique orders within each block in non-increasing order of right degree.

By \autoref{lem:cluster-block-order}, this merge preserves the relative order within every clique. The resulting ordering has modulator positions $P_\alpha$, modulator right degrees $D_\alpha$, and groups of clique vertices described by $h_{j,r}$ and $q_{j,r}$. Its cost is therefore exactly the IQP objective value.

Conversely, consider a minimum-cost ordering whose restriction to $M$ is $\sigma_M$. By \autoref{lem:cluster-block-order}, we may assume that it satisfies the stated ordering properties. Setting $x_{i,f}$ to the number of cliques of class $A_i$ following profile $f$ gives a feasible IQP solution with the same cost.

Thus, every feasible IQP solution yields an ordering of the same cost, and a minimum-cost ordering yields a feasible IQP solution of the same cost. The optimum values are equal.
\qed
\end{proof}
\subsection{Proof of \autoref{thm:clusterruntime}}
\begin{proof}
Cluster modulator $M$ of a graph can be found in time $O^*(1.811^k)$ \cite{tsur2021faster}. 
For each of the $k!$ permutations of $M$, construct and solve the corresponding reduced IQP. By \autoref{t3}, its optimum equals the minimum cost among orderings extending that permutation. We reconstruct an ordering from each solution and return one of minimum cost. The FPT running time is a function of $k+d$ as follows:
To exploit the structure of our constraints, substitute the expressions for $P_\alpha$, $D_\alpha$, $h_{j,r}$, and $q_{j,r}$ into the objective. These quantities are affine functions of the profile-count variables, so the resulting objective remains quadratic.
The reduced IQP has only the variables $x_{i,f}$ and the constraints
$\sum_{f\in\mathcal{F}_i}x_{i,f}=|A_i|
    \text{ for all }i\in[t], \quad 
    x_{i,f}\in\mathbb{Z}_{\geq0}
    \text{ for all }i\in[t],\ f\in\mathcal{F}_i.
$
Its constraint matrix is totally unimodular: each variable belongs to exactly one class-sum equation, and adding nonnegativity constraints or replacing an equality by two inequalities preserves total unimodularity.
After doubling the objective and dropping its constant term, write it as $x^\top Hx+c^\top x$, where $H$ is a symmetric integer matrix. We claim that $\|H\|_{\max}=O((k+d)d^2)$. Indeed, each coefficient in the linear parts of $P_\alpha$ and $D_\alpha$ is at most $d$, so their products contribute $O(kd^2)$ to any quadratic coefficient.

For the contribution from clique vertices, a fixed profile satisfies $\sum_{j,r}y_{i,f,j,r}=d_i\leq d$. Moreover, its coefficient in any expression $q_{j,r}$ is at most $d$, since it counts vertices of a single clique preceding that group. As $r\leq R$, the terms involving $h_{j,r}q_{j,r}$ and $h_{j,r}^2$ contribute $O(Rd^2)$ to any quadratic coefficient. Together, these bounds establish the claim.

Let $N=\sum_{i=1}^{t}|\mathcal{F}_i|$ be the number of profile-count variables, and define $B=\left[\binom{2^k+d}{d}-1\right](k+1)^d$. Then $N\leq B$.
By Corollary~1.2 of \cite{ari2026curvature}, an IQP with $N$ variables, the running time for each permutation is,
$
    \bigl[B(k+d)d^2\bigr]^{O(B)}n^{O(1)}
    =\bigl[B(k+d)\bigr]^{O(B)}n^{O(1)}.
$
\qed
\end{proof}
\section{Proofs missing in Section~\ref{sec:split}}
\subsection{Proof of \autoref{lem:unit-degree-exchange}}
\begin{proof}
Write $rd(v)=d$. Every vertex strictly between $u$ and $v$
has right degree $d+1$ or $d$. Let $H$ and $L$ be the sets
of intervening vertices of these respective degrees.
The sets $\{u\}\cup H$ and $L\cup\{v\}$ are independent.

Suppose that some $h\in H$ is adjacent to $v$. Permute $H$ so that $h$ is
last, and $L\cup\{v\}$ so that $v$ is first. These permutations preserve the
cost and make $h$ and $v$ consecutive. Swapping this consecutive pair also
preserves the cost, since they are adjacent and their right degrees differ by
one. The swap increases the right degree of $v$ to $d+1$, while that of $u$
remains $d+1$. The resulting optimal ordering has adjacent vertices $u$ and
$v$ of equal right degree, a contradiction.

Similarly, suppose that some $\ell\in L$ is adjacent to $u$. Permute
$\{u\}\cup H$ so that $u$ is last, and $L$ so that $\ell$ is first. Swapping
the now-consecutive vertices $u$ and $\ell$ preserves the cost and decreases
the right degree of $u$ to $d$. The right degree of $v$ remains $d$, giving
the same contradiction.

Thus neither endpoint has a neighbor strictly between them. Set
$D=\phi(v)-\phi(u)$. The edge $uv$ retains the same contribution when the
endpoints are swapped; the other incident edges change the cost by
$D\bigl(rd(u)-1-rd(v)\bigr)=0$. Hence the swap preserves the cost.
\qed
\end{proof}
\subsection{Proof of \autoref{lem:insertion-estimate}}
\begin{proof}
Let $w_1,\ldots,w_{\ell}$ be the vertices at positions
$\phi(x),\ldots,\phi(y)-1$, in their order of appearance.
Thus $w_1=x$.

Construct an ordering $\lambda$ by moving $y$ immediately
before $x$ and shifting each $w_i$ one position to the
right. Every edge from $y$ to a vertex originally after $y$ decreases its contribution by $\ell$.
For each $w_i$, shifting it one position to the right
increases the contribution of each of its right edges
by one, except for $w_iy$, if that edge exists.
The contribution of $w_iy$ actually changes by $-(i-1)$,
rather than $+1$, so this edge requires a correction
of $-i$. All other edges are unaffected. Therefore,
\[
\mu(\lambda)-\mu(\phi)
=
\sum_{i=1}^{\ell}rd_{\phi}(w_i)
-\ell\,rd_{\phi}(y)
-\sum_{\substack{1\le i\le\ell\\w_iy\in E}}i.
\]

Right degrees are non-increasing along $\phi$, and hence $rd_{\phi}(w_i)\le rd_{\phi}(x)$ for every $i$. Moreover, the indices of any $r$ neighbors of $y$
are distinct positive integers, so their sum is at least
$r(r+1)/2$. Consequently,
\[
\mu(\lambda)-\mu(\phi)
\le
\ell\bigl(rd_{\phi}(x)-rd_{\phi}(y)\bigr)
-\frac{r(r+1)}2.
\]
Since $\phi$ is optimal, the left-hand side is
nonnegative, proving the inequality.
\qed
\end{proof}
\subsection{Proof of \autoref{lem:g-v}}
\begin{proof}
For any ordering $\lambda$ of $G-v$, placing $v$ first
and shifting $\lambda$ by one produces an ordering of cost $\deg_G(v)+|E(G-v)|+\mu_{G-v}(\lambda)$. The first term accounts for the edges incident to $v$;
the second accounts for shifting the edges of $G-v$.
Both terms are independent of $\lambda$. If an ordering
of $G-v$ had lower cost than $\psi$, this construction
would improve on $\phi$, a contradiction.
\qed
\end{proof}
\subsection{Proof of \autoref{lem:zero-degree-placement}}
\begin{proof}
Suppose that $u\in I$ has right degree zero and precedes
$w\in Q\cup M$. By right-degree monotonicity, all vertices
from $u$ to $w$ have right degree zero. Swapping $u$ and
$w$ within this equal-degree block preserves the cost
and strictly increases $P$, a contradiction.
\qed
\end{proof}
\subsection{Proof of \autoref{lem:surplus-monotonicity}}
\begin{proof}
Let $u,v\in I$ with $\phi(u)<\phi(v)$, and let $p$ be the number of clique vertices between them. Then $q_{\phi}(u)=q_{\phi}(v)+p$.
If $p=0$, monotonicity of right degrees gives $rd_{\phi}(u)\ge rd_{\phi}(v)$.
Since $q_{\phi}(u)=q_{\phi}(v)$, it follows that $s_{\phi}(u)\ge s_{\phi}(v)$.

Suppose that $p\ge 1$, and let $x_1,\ldots,x_p$ be the clique vertices between
$u$ and $v$, listed in their order of appearance. By monotonicity,
$rd_{\phi}(u)\ge rd_{\phi}(x_1)$. If equality held, all vertices from $u$ to
$x_1$ would belong to an equal-degree block. Swapping $u$ and $x_1$ would
therefore preserve the cost while strictly increasing the potential $P(\phi)$, a
contradiction. Hence, $rd_{\phi}(u)\ge rd_{\phi}(x_1)+1$.

Since the vertices $x_1,\ldots,x_p$ form a clique, their right degrees decrease strictly. Thus, $rd_{\phi}(x_1)\ge rd_{\phi}(x_p)+p-1$. Moreover, $rd_{\phi}(x_p)\ge rd_{\phi}(v)$ by monotonicity. Combining these inequalities gives $rd_{\phi}(u)\ge rd_{\phi}(v)+p$. Consequently,
\[
s_{\phi}(u)
=rd_{\phi}(u)-q_{\phi}(u)
\ge rd_{\phi}(v)+p-\bigl(q_{\phi}(v)+p\bigr)
=s_{\phi}(v).
\]
\qed
\end{proof}
\subsection{Proof of \autoref{lem:zero-surplus}}
\begin{proof}
If $q(u)=0$, then $s(u)=rd(u)\ge0$. Otherwise, let $y\in Q$ be the first clique vertex after $u$. Monotonicity gives $rd(u)\ge rd(y)$. Equality would allow
a cost-preserving swap of $u$ and $y$ that increases $P$,
so $rd(u)\ge rd(y)+1$. Since $y$ is adjacent to the other $q(u)-1$ clique vertices after $u$, $rd(u)\ge rd(y)+1\ge q(u)$.
Thus $s(u)\ge0$.

Now suppose that $s(u)=0$ and that $u$ has a clique
neighbor after it. Let $x\in Q$ be the first such neighbor,
and let $t$ count the intervening clique vertices.
By \autoref{lem:first-clique-neighbor-drop},
\begin{equation}
\label{eq:zero-surplus-drop}
rd(x)\le rd(u)-t-2=q(u)-t-2=q(x)-1.
\end{equation}
This contradicts $rd(x)\ge q(x)$, since $x\in Q$.
Hence $rc(u)=0$. All right neighbors of $u$ then belong
to $M$, giving 
$q(u)=rd(u)\le k$ (note that $q(u)=rd(u)$ by assumption $s(u)=0$). Together with
non-negativity of surplus, this proves the final assertion.
\qed
\end{proof}
\subsection{Proof of \autoref{lem:equal-surplus-clique-bound}}
\begin{proof}
If $u$ and $z$ belong to the same independent group,
there is no clique vertex between them, and the claim
follows.

Otherwise, let $u'\in I$ be the last vertex of the independent group
containing $u$. Since $u'$ occurs between $u$ and $z$, surplus monotonicity
gives $s(u)\ge s(u')\ge s(z)$, and therefore $s(u')=s$.

Suppose that a clique neighbor of $u'$ occurs before $z$. Let $x\in Q$ be the
first clique neighbor of $u'$ after $u'$, and let $b\in I$ be the first
independent set vertex after $x$. The vertex $b$ exists because $z$ occurs after
$x$, and $\phi(b)\le\phi(z)$. By \autoref{lem:surplus}, $s(b)<s(u')$. Surplus
monotonicity then gives $s(z)\le s(b)<s(u')=s$, a contradiction.

Thus every clique vertex between $u'$ and $z$ is a non-neighbor of $u'$. Since
$I$ is independent, $rd(u')-rc(u')$ counts the right modulator neighbors of
$u'$ and is at most $k$. Consequently, $q(u')-rc(u') = rd(u')-rc(u')-s(u') \le
k-s$. The left-hand side counts the clique non-neighbors of $u'$ to its right.
Hence at most $k-s$ clique vertices occur between $u'$ and $z$.

Finally, no clique vertex occurs between $u$ and $u'$, because they belong to
the same independent group. The same bound therefore holds between $u$ and
$z$.
\qed
\end{proof}
\subsection{Proof of \autoref{lem:after-threshold-bound}}
\begin{proof}
Let $V_2=\{v\in V:\phi(v)>T\}$ and $G_2=G[V_2]$. Since $I$ is independent, the set $(Q\cup M)\cap V_2$
is a vertex cover of $G_2$. There are fewer than $2k$
clique vertices and at most $k$ modulator vertices
after $T$, so this vertex cover has size at most $3k$.

By repeated application of \autoref{lem:g-v}, the ordering $\phi_2$ of $G_2$
given by $\phi_2(v)=\phi(v)-T$ for $v\in V_2$ is optimal. Let $\tau$ be the minimum vertex-cover size.
Zhang and Cao prove that every optimal ordering of a graph with $\tau\ge2$
has at most $2\tau\log_2\tau$ vertices of positive right degree
\cite[Theorem~1.1]{zhang2026minimum}. 
Since $\tau\le3k$, in all cases the number is at most
$\beta(k)=\lceil6k\log_2(3k)\rceil=O(k\log(k+1))$.

For every $v\in V_2$, all vertices after $v$ in $\phi$ also belong to $V_2$,
and their relative order is unchanged. Hence, $rd_{\phi_2}(v)=rd_{\phi}(v)$.
This proves the positive-right-degree bound after $T$. The final assertion
follows because positive surplus implies positive right degree.
\qed
\end{proof}
\subsection{Details of the split-modulator algorithm}
\label{app:split-modulator-xp}

\paragraph{Cost decomposition.}
Fix an enumerated partition $(V_1,V_2)$ and let $T=|V_1|$.
For orderings $\rho_1$ of $V_1$ and $\rho_2$ of $V_2$, define
their concatenation by
\[
\rho(v)=
\begin{cases}
\rho_1(v), & v\in V_1,\\
T+\rho_2(v), & v\in V_2.
\end{cases}
\]
Every edge between the parts contributes the position of its endpoint
in $V_1$. Hence all such contributions are captured by
$w(v)=|N(v)\cap V_2|$ in the weighted cost of $G_1$.
Each edge within $V_2$ contributes its cost under $\rho_2$ plus $T$.
Together with the edges within $V_1$, this proves
\eqref{eq:split-decomposition-cost} for arbitrary orderings of the
two parts. Since $T|E(G_2)|$ is constant for the fixed partition,
independently optimal suborderings give an optimum among all orderings
placing $V_1$ before $V_2$.

\paragraph{Partition enumeration.}
Write $n_1=|Q|$ and $n_2=|I|$. There are at most
\begin{equation}
\label{eq:split-partition-count}
2^k
\left(\sum_{i=0}^{\min\{\alpha,n_2\}}\binom{n_2}{i}\right)
\left(\sum_{j=0}^{\min\{2k,n_1\}}\binom{n_1}{j}\right)
=2^k n^{O(k^3)}
\end{equation}
choices of $(I_1,Q_2,M_1)$ for $k\ge1$.
By \autoref{lem:positive-surplus-bound} and the threshold bounds,
at least one choice agrees with the partition of an optimal ordering.
For each candidate, deleting $I_1\cup M_1$ from $G_1$ leaves the
clique $Q\setminus Q_2$, so its clique modulator has size at most
$\alpha+k$. Deleting $Q_2\cup(M\setminus M_1)$ from $G_2$ leaves
the independent set $I\setminus I_1$, so its vertex cover has size
at most $3k$.

\paragraph{Solving the second instance.}
The bound used in \autoref{lem:after-threshold-bound} implies that
every optimal ordering of $G_2$ has at most
$\beta=\lceil6k\log_2(3k)\rceil=O(k\log(k+1))$ vertices of
positive right degree~\cite[Theorem~1.1]{zhang2026minimum}.
Right degrees are non-increasing in an optimal ordering, so these
vertices form a prefix. Enumerate every set $S\subseteq V_2$ with
$|S|\le\beta$ and every ordering of $S$. Retain a choice only if
$V_2\setminus S$ is independent, and append that complement in
arbitrary order.

All appended vertices have right degree zero. Their relative order
does not change the cost: edges from $S$ to the complement contribute
the positions of their endpoints in $S$, and no edge lies within the
complement. The enumeration includes the positive-right-degree prefix
of an optimal ordering, with its order. Thus a minimum-cost candidate
is optimal. The number of candidates is at most
\[
\sum_{i=0}^{\min\{\beta,|V_2|\}}
\binom{|V_2|}{i}\,i!=n^{O(\beta)}.
\]
This includes the empty prefix when $G_2$ has no edges. Constructing
and evaluating candidates takes polynomial time per candidate.

\paragraph{Correctness and total running time.}
For $k=0$, the input is a split graph and can be solved in polynomial
time. For $k\ge1$, consider an enumerated partition that agrees with
an optimal ordering of $G$. The cost decomposition shows that
replacing its two suborderings by independently optimal ones cannot
increase the cost. Their concatenation is therefore optimal for $G$.
Every other candidate is also an ordering of $G$, so returning a
minimum-cost candidate over all partitions gives an optimal solution.

For each partition, the weighted clique-modulator algorithm of
\autoref{subsection:wtclique} runs in XP time with parameter at most
$\alpha+k=O(k^3)$. Solving $G_2$ takes
$n^{O(k\log(k+1))}$ time. Multiplying these bounds by the partition
count in \eqref{eq:split-partition-count} yields total running time
$f(k)n^{g(k)}$ for computable functions $f$ and $g$, proving
\autoref{thm:split-modulator-xp}.

\end{document}